\documentclass[11pt]{article}
\usepackage{graphicx}
\usepackage{amsmath}
\usepackage{amssymb}
\usepackage{amsxtra}

\title{Anomalous Isotopes In Dark Atoms models}
\author{M.I. Baliño$^{1}$, J.Mwilima$^{1}$, D. O. Sopin$^{1,2}$, M. Yu. Khlopov$^{3}$, \\
$^1$ National Research Nuclear University MEPhI 115409 Moscow, Russia\\
$^2$ Research Institute of Physics, Southern Federal University 344090
\\Rostov on Don, Russia\\ 
$^3$ Virtual Institute of Astroparticle physics 75018 Paris, France\\
\\balinomagela@gmail.com\\
joshuamwilima02@gmail.com\\
khlopov@apc.univ-paris7.fr\\
sopin@sfedu.ru}

\begin{document}
\maketitle

\begin{abstract}
In this work, we study some aspects of the dark atom model. We consider a finite-size nucleus to find the wave functions of the bound state of a stable particle with a charge of $-2n$ and helium-4 $^4He^{++}$. Then we address the problem of calculating the abundance of anomalous isotopes arising from the capture of helium nuclei by dark atoms during Big Bang nucleosynthesis. We use an analogy with the proton–neutron capture process to calculate the reaction cross section and thus determine the concentration of OBe nuclei.
\end{abstract}

\section{Introduction}\label{s:intro}

Dark matter constitutes about 26\% of the total energy density of the Universe, while its nature remains one of the key unsolved problems in modern cosmology \cite{Planck2018}. Convincing evidence for its existence is manifested in the behavior of galaxies, gravitational lensing, and the anisotropy of the cosmic microwave background radiation. These observations indicate that dark matter is non-baryonic in nature. Various candidates for its composition are discussed, including weakly interacting massive particles (WIMPs), supersymmetric particles (SUSY), and other similar possibilities. The absence of positive detection results for WIMPs and supersymmetric particles at the Large Hadron Collider (LHC) makes it interesting to consider alternative candidates for dark matter.

A promising explanation is provided by the concept of \emph{dark atoms}. 
In this scenario, a new heavy stable particle $X$ with electric charge $-Z_X=-2n$ (where $n$ is a natural number) binds with $n$ helium nuclei ($^4$He) through the electromagnetic Coulomb interaction, forming neutral bound states.  
These composite systems, often referred to as $XHe$, represent a viable candidate for dark matter \cite{Khlopov2021,Beylin2024}. The model can explain the paradoxes arising in the search for dark matter particles in underground experiments.

In the simplest case $n=1$ ($Z_X=2$), the system forms an $OHe$ atom consisting of one heavy particle $X$ with charge $-2$ (denoted $O^{--}$), and one $^4$He nucleus with charge $+2$. The properties of such bound states are determined by their internal structure, the analysis of which is one of the main objectives of this study.
%We consider the quantum description of the dark atom in the case of OHe, taking into account the finite size of the helium nucleus. The model of the alternative kind of the hydrogen atom (AKHA) \cite{2}, was also considered.

$OHe$ atoms, created during Big Bang Nucleosynthesis in reaction
\[
O^{--} + \,^4\mathrm{He}^{++} \;\;\longrightarrow\;\; OHe,
\]
may interact with ordinary nuclei  and thereby influence the chemical and cosmological evolution of the early Universe. In particular, the capture of $^4$He nuclei by $OHe$ can lead to the formation of anomalous isotopes such as $OBe$, whose abundance and physical consequences are the subject of this work. 

This paper is organised as follows: in section \ref{s:Schrodinger} the Schrodinger equation is solved to find the wave functions of dark atoms. Then, in section \ref{s:Interaction}, the probability and cosmological consequences of helium capture by dark atoms are estimated. We analyse the corresponding cross-sections estimated by analogy with the radiative capture of neutrons by protons and evaluate the resulting abundance of anomalous isotopes. The obtained results are briefly discussed in section \ref{s:discussion} and in the Conclusion.

\section{Solution of the Schrodinger equation \label{s:Schrodinger}}

%\subsection{Singular solution}
%It is known that there exist two classes of solutions to the Schrodinger equation describing motion in a centrally symmetric field \cite{2,3}. These two solutions are characterized by different behaviors at small values of radial variable:
%\begin{itemize}
%    \item $R(r) \approx r^l$ - regular solution;
%    \item $ R(r) \approx\frac{1}{r^{l+1}}$ - singular solution.
%\end{itemize}
%In the case of continuous spectrum solutions, the second one is taken into account when finding the scattering phase. However, the solutions of the discrete spectrum are determined by the boundary condition at the origin, which does not allow consideration of an irregular solution. It was pointed out in \cite{2} that the only reason for using it and rejecting the second solution is the diverging normalization integral \(\int_{0}^{\infty}|R(r)|^2r^2 dr\). However, the author demonstrated that by taking into account the finite size of the nucleon and matching the internal solution with the external singular solution at the nucleon boundary, this normalization issue can be resolved. It is possible for a special class of nuclear potentials.

%А что дальше? Разлоение в ряд требует аналитичности, на что Окс ответил в 1 статье. Нефизичность относительно l закрыто требованием основного состояния.
%Следовательно, у нас нет рабочих аргументов против Окса. А раз так, то смысл этого раздела пропадает?

The inner structure of the dark atom should be similar to that assumed in Thomson's plum pudding model. Indeed, in most cases, the Bohr radius (in natural units: $\hbar=c=1$) $r_B=(Z_N Z_X \alpha m_N)^{-1}$ of the nucleus $N$ in the shell of such bound states is smaller than the nuclear charge radius $r_N$. Therefore, to describe the structure of the dark atom, it is necessary to consider the eigenvalue problem for the Schrodinger equation with a piecewise potential. The inner part should describe the charge distribution inside the nucleus, while the external part must coincide with the Coulomb potential. 

The simplest choice is to consider a nucleus as a uniformly charged sphere. This assumption leads to an oscillatory potential 
\begin{equation}
    V(\rho)=
    \begin{cases}
        \cfrac{1}{a}\left(3-\cfrac{\rho^2}{a^2}\right), \rho<a;
        \\
        \cfrac{2}{\rho},\rho>a,
    \end{cases}
\end{equation}
where $\rho=r/r_B$, $a=r_N/r_B$. It describes the intersection of the heavy multicharged particle $X^{-2n}$ and an ordinary, relatively light isotope $N$ in the center of mass system. Other charge distributions require additional parameters, the values of which are determined by experiments. On the one hand, a more accurate spherically symmetric potential should provide only an insignificant correction for most combinations of particles. On the other hand, the deformation of the nuclei caused by the presence of a multicharged core eliminates the expected gain in accuracy. However, in several special cases, the use of spherically symmetric potentials may lead to unreliable predictions. In particular, to describe the structure of the anomalous isotope $OBe^{++}$, it is necessary to make  more accurate calculations.

\begin{table}
    \centering
    \begin{tabular}{|c|c|c|c|c|c|c|}
         \hline
            n & 1 & 2 & 3 & 4 & 5 \\
        \hline
              $r_N$, \mbox{Fm} & \multicolumn{5}{c|}{1.678 \cite{Antognini:2021icf,Krauth2021}} \\
             $r_B, 10^{-3}\,\mbox{MeV}^{-1}$ & 9.19 & 4.60 & 3.06 & 2.30 & 1.84 \\
             $a$ & 0.913 & 1.826 & 2.739 & 3.652 & 4.566 \\
        \hline
        \hline
            $E^{\text{Coulomb}}_{XN},\,\mbox{MeV}$ & 1.588 & 6.352 & 14.291 & 25.406 & 39.698 \\
            $E^{\text{Piecewise}}_{XN},\,\mbox{MeV}$ & 1.256 & 3.891 & 7.130 & 10.708 & 14.506 \\
        \hline
    \end{tabular}
    \caption{Properties of $X^{-2n}-^4$He bound states}
    \label{tab:4HeErparam}
\end{table}

\begin{figure}
    \begin{minipage}{0.49\linewidth}
    \centering{\includegraphics[width = 1\linewidth]{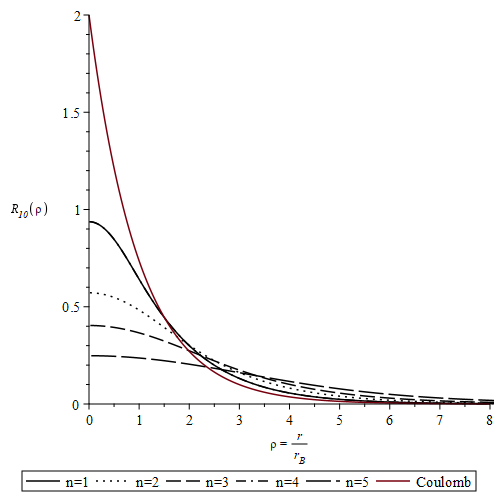}} 
    \end{minipage}
    \begin{minipage}{0.49\linewidth}
    \centering{\includegraphics[width = 1\linewidth]{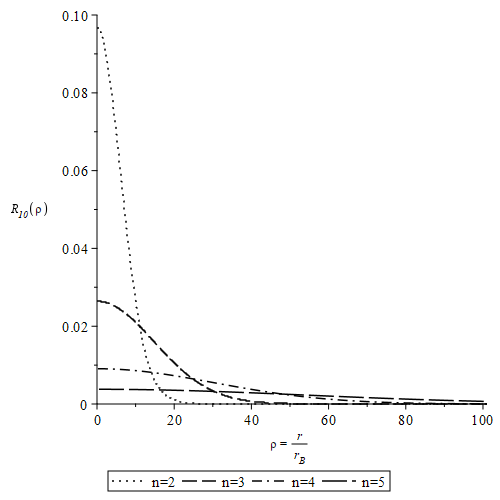}} 
    \end{minipage}
    \caption{The physical radial functions $R(\rho)=P(\rho)/\rho$ for $X^{-2n}-^4$He (left panel) and electrically neutral $X-N$ bound states.}
    \label{Fig:WaveFun}
\end{figure}

The radial Schrodinger equation in considered unites is
\begin{equation}
    \partial_{\rho}^2P(\rho)+\left(\varepsilon-\cfrac{l(l+1)}{\rho^2}+V(\rho)\right)P(\rho)=0,
\end{equation}
where $\varepsilon=2 m_N r_B^2 E_{XN}$, $P(r)=rR(r)$. This eigenvalue problem solution for different charges of the heavy core $X$ is presented in Table \ref{tab:4HeErparam}. The obtained values of the ground state binding energy are significantly smaller than the Bohr-like estimate $E^{\text{Coulomb}}_{XN}=(2 m_N r_B^2)^{-1}$ predicts. The corresponding eigenfunctions in the form of the physical radial functions $R(\rho)=P(\rho)/\rho$ for $^4$He nucleus are shown in the left panel of Figure \ref{Fig:WaveFun}. Although the value at the origin decreases with increasing core charges, the probability of finding the heavy particle $X$ inside the nucleus grows due to the change of the Bohr radius. The similar dependence may be found for the neutral states (see the right panel of Figure \ref{Fig:WaveFun}). 

%The rates of reactions in which dark atoms participate depends on the properties of these bound state. For instance, to find the cross section of dark recombination we need to solve the eigenvalues and eigenfunctions problem for Schrodinger equation with a piecewise potential. The Bohr radius of 

\section{Interaction with nuclei \label{s:Interaction}}
\subsection{Rates of reactions}
To produce the correct estimation of anomalous isotope concentration, it is necessary to consider at least two new reactions: dark atom recombination (radiative capture of the first helium) and the capturing of additional light nucleus. The ratio of the first of them can be found with the analogy of ordinary hydrogen recombination. The rescaled semiclassical Kramer's formula \cite{Kotelnikov:2018woh} was used:
\begin{equation}
    \left<\sigma v\right>_{\text{rec}}=\cfrac{32}{3}\sqrt{\cfrac{\pi}{3}} \cfrac{Z_N}{Z_X} \cfrac{1}{m_N^4 r_B^2} \left(\cfrac{E_{X-N}}{T}\right)^{\frac{1}{2}}\left(\ln{\left(\cfrac{E_{X-N}}{T}\right)}+\gamma\right),
    \label{recRate}
\end{equation}
where $\gamma =  0.5772$ is an Euler constant. It approximately takes into account the transitions on excited states and therefore the estimation should be more accurate than with Stobbe formula. In case of ordinary recombination, the exact rate is 3.2\% less than calculated with \eqref{recRate}. However, in the processes that involve the dark atoms, the structure of the finite-size nucleus should be significant. It leads to a higher error and makes it possible only to get a quite accurate estimation.

%In the considered model, $OHe$ atoms can interact
The rate of the $OHe$ atom interaction with light nuclei such as $^4$He can be estimated by analogy with the radiative capture of neutrons by protons, taking into account:
\begin{itemize}
  \item the absence of $M1$ transition (orbital angular momentum conservation),
  \item suppression of the $E1$ transition for the $OHe$ system.
\end{itemize}
Since $OHe$ is isoscalar, the isovector $E1$ transition is only possible due to isospin violation, parameterised by $f \sim 10^{-3}$. The resulting capture rate is given by \cite{Khlopov2010}:

\begin{equation}
\left<\sigma v\right> = \frac{f \alpha}{m_p^2} 
\cdot \sqrt{\frac{3}{2}} 
\left(\frac{Z}{A}\right)^{2} 
\cdot \frac{T}{\sqrt{A m_p E}} \, .
\end{equation}

where $A$ and $Z$ are the atomic mass and charge numbers, $E$ is the binding energy of the state, and $T$ is the plasma temperature.  
For $^4$He ($A=4$, $Z=2$), $E_{X-N} \approx 1.6$ MeV. At $T \sim 100$ keV the cross-section is of order $\sim 10^{-36}\ \text{cm}^2$.

The rate of $OHe$ photodestruction can be found with detailed equilibrium:
\begin{equation}
    \cfrac{\left<\sigma v\right>_{\text{rec}}}{\left<\sigma v\right>_{\gamma}}=\left<\cfrac{2 E^2_{\gamma}}{p^2_N}\right>\approx\cfrac{2 E^2_{X-N}}{m^2_N}\sqrt{\cfrac{2m_N}{\pi T}},
\end{equation}
where the approximation $E_{\gamma}\approx E_{X-N}$ was used. The right side of the equation is averaged over the Maxwellian distribution.

\subsection{Numerical Estimation of $OBe$ Abundance with LINX}

To estimate the abundance of anomalous isotopes produced after the interaction of $OHe$ with ordinary nuclei, 
we employed the nucleosynthesis code \texttt{LINX} \cite{LINX}, which uses methods and tables from \cite{nudecBSM1,nudecBSM2,PRIMAT,PRyM}.
This numerical framework is designed for modeling nuclear reactions under the conditions of Big Bang Nucleosynthesis (BBN). 
For the purposes of this study, we have introduced additional particle species ($OHe$ and $OBe$) 
and implementing new reaction channels describing the radiative capture of two $^4$He by new charged particle $O^{--}$. %$OHe$ atoms.  

The reaction cross-sections, described in Section~\ref{s:intro}, were incorporated into the network of reactions, allowing us to evolve the abundances consistently with cosmological parameters. However, the program requires to include the relative concentration of photons with high enough energy $Y_{\gamma}(T)=n_{\gamma}/n_b$ by hands in case of radiative reactions. Therefore, it is necessary to find
\begin{equation}
     Y_{\gamma}\cfrac{\left<\sigma v\right>_{\gamma}}{\left<\sigma v\right>_{\text{rec}}}=\cfrac{\delta \pi^{3/2}}{2 \eta \zeta(3)}m_N^{3/2}E^{\xi-2}_{X-N}T^{-\xi+1/2}\exp\left(\frac{(\kappa-1)E_{X-N}}{T}\right),
\end{equation}
where $\delta$, $\xi$ and $\kappa$ is obtained by approximation of
\begin{equation}
    Y_{\gamma}=\frac{1}{\eta}\cfrac{\pi^2}{2\,\zeta(3)\,T^3}\int^{\infty}_{E_{X-N}}\cfrac{E^2}{\exp\left(\frac{E}{T}\right)-1}dE,
\end{equation}
$\eta=6.04\cdot10^{-10}$ is the baryon-to-photon ratio and $\zeta(3)\approx1.202$ is the value of zeta function. The photodestruction $OBe+\gamma\rightarrow OHe+He$ becomes possible only at low energies, when the excess of $XHe$ is generated. Also it requires at least the same energy of photon. Therefore, we can neglect it: $\left<\sigma v\right>_{\gamma}/\left<\sigma v\right>_{\text{rec}}\approx0$.

\begin{figure}[!ht]
\centering
\includegraphics[scale=0.5]{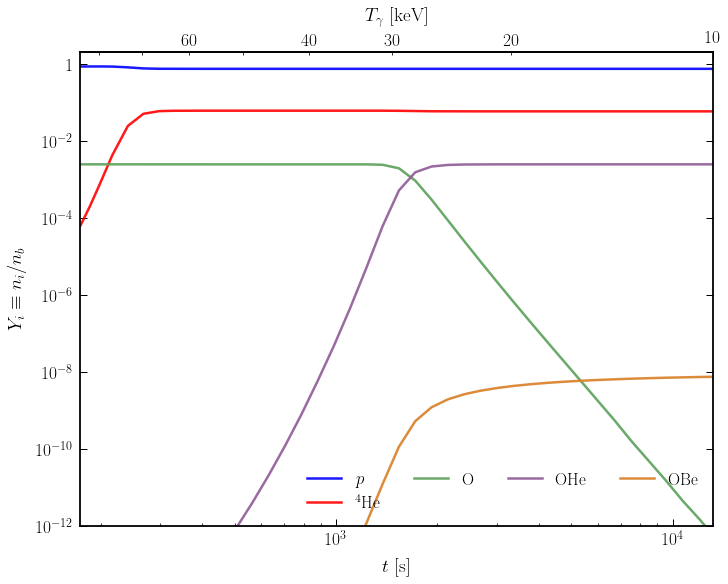} 
 \caption{Concentrations of bound states during nucleosynthesis
 }
 \label{Fig:O-OHe-OBe}
\end{figure}

Finally, with the assumption that initially all of the dark matter density is provided by $O^{--}$ ($\frac{\rho_{DM}}{\rho_B} \approx 5.36$) the Fig. \ref{Fig:O-OHe-OBe} can be calculated. The relative concentrations of dark matter particles at the end of nucleosynthesis ($T_{\text{end}}\approx5$ keV) are
\begin{equation}
    Y_{O^{--}}\approx1.2\cdot10^{-18},\qquad Y_{OHe}\approx2.5\cdot10^{-3},\qquad Y_{OBe}\approx 8.5\cdot 10^{-9}
\end{equation}
Almost all charged lepton-like particles recombine with helium. Moreover, there is the significant overproduction of anomalous isotopes $OBe^{++}$. The process of its formation freezes out shortly after the dark atom neutralization.

%which corresponds to an initial mass fraction $Y_{OHe} \approx 0.0025$.  

%\textit{(Here Danila should explain in more detail: not all photons in the Universe are capable of destroying dark atoms..... )}

%\subsection{Simulation Results}
%According to the modified \texttt{LINX} code, the final abundance of anomalous isotopes is strongly suppressed:
%\begin{equation}
%Y_{OBe}^{\text{final}} \approx jjj
%\end{equation}

\section{Discussion \label{s:discussion}}

The obtained result is only a preliminary estimation. To find the realistic concentrations of anomalous isotopes in the dark atom model, a more comprehensive treatment of nuclear processes is necessary.

Proton capture is expected to be strongly suppressed due to the excess of high-energy photons. Using the Saha-like formula, it is possible to estimate the temperature at which dark recombination with isotopes of hydrogen becomes possible. It is a few keV. At this time the concentration of free negatively charged particles should be negligible and, therefore, there is no  production of $OH$ bound states. Nevertheless, late proton capture is dangerous.  Although $^5$Li cannot be stabilized in a dark atom shell solely by the suppression of Coulomb repulsion, the synthesis of other lithium isotopes may be catalyzed  \cite{Akhmedov:2024fpw}.  The overproduction of primordial metals and anomalous isotopes is constitutes the central challenge for dark atom nucleosynthesis scenarios. 

A potential resolution of this problem for the case of doubly charged particles ($n=1$) may be found by considering the reactions $XN_1+N_2\rightarrow XN_3+N_4+...$. The Fig. \ref{Fig:O-OHe-OBe} shows that anomalous isotope production freezes out at relatively high temperatures. The interaction with primordial plasma may lead to the destruction of anomalous bound states at late nucleosynthesis stages. However, this requires much more careful analysis.

Generally, the same problems arise for all values of heavy core charges.   However, there are several qualitative changes. First of all, the increased binding energy of dark ions prevents the destruction of the nuclear shell. Moreover, the Coulomb repulsion should be significantly suppressed. It allows for the consideration of nuclei that are unstable in the free state. The lifetime of the proton-rich isotopes of beryllium, boron and carbon may be extended due to the suppression of proton emission in the shell of the dark atom. This also should catalyze the formation of $XC$ bound states  in a 3-$\alpha$ process, analogous to the one occurring during stellar nucleosynthesis.

For higher-charge heavy cores, dark atom recombination becomes a multi-stage process. Therefore, the probability of hydrogen capture should increase. This may open new channels of dark atom recombination. In particular, for $n\geq5$, the capture of protons at the initial stage of this process should be a dominant process. Therefore, the enhanced overproduction of odd-charged dark ions is expected. On the other hand, numerical estimation of binding energies shows that for $n\geq4$, the formation of the intermediate bound state $(XHe)p$ becomes possible. The main feature of this configuration is the absence of nuclear fusion within the shell of dark ion, which could significantly alter the subsequent nucleosynthetic pathway..

Finally, the large number of stages leads to the prolongation of the overall dark atom recombination timescale. Consequently, a sufficient concentration of neutral states may only be formed at low temperatures, when the production of anomalous isotopes should freeze out. 

\section{Conclusion}

Heavy, stable, multicharged particles $X^{-2n}$, predicted in several extensions of the Standard Model, should bind with light primordial nuclei during Big Bang nucleosynthesis to form the neutral dark atoms. However, a more careful consideration of this process reveals some problems of the dark atom scenario. In particular, the simple estimation of particle abundances indicates a significant overproduction of anomalous isotopes. To provide the realistic estimation of ordinary and dark matter particle concentrations, it is necessary to include the additional reactions.

\section*{Acknowledgements}
The work of D.O.S. was performed in Southern Federal University with financial support of grant of Russian Science Foundation № 25-07-IF.

%% The bibliography section

\end{document}